\documentclass[preprint2, twocolumn]{aastex631}

\usepackage{graphicx}	
\usepackage{amsmath}	

\begin{document}

\title{Estimate Sonic Mach Number in the Interstellar Medium with Convolutional Neural Network}

\author{Tyler Schmaltz}
\affiliation{Department of Physics, University of Wisconsin-Madison, Madison, WI, 53706, USA}
\affiliation{Department of Astronomy, University of Wisconsin-Madison, Madison, WI, 53706, USA}

\author[0000-0002-8455-0805]{Yue Hu*}
\affiliation{Institute for Advanced Study, 1 Einstein Drive, Princeton, NJ 08540, USA}

\author{Alex Lazarian}
\affiliation{Department of Astronomy, University of Wisconsin-Madison, Madison, WI 53706, USA}

\email{yuehu@ias.edu; alazarian@facstaff.wisc.edu; *NASA Hubble Fellow}



\begin{abstract}
Understanding the role of turbulence in shaping the interstellar medium (ISM) is crucial for studying star formation, molecular cloud evolution, and cosmic ray propagation. Central to this is the measurement of the sonic Mach number ($M_s$), which quantifies the ratio of turbulent velocity to the sound speed. In this work, we introduce a convolutional neural network (CNN)-based approach for estimating $M_s$ directly from spectroscopic observations. The approach leverages the physical correlation between increasing $M_s$ and the shock-induced small-scale fluctuations that alter the morphological features in intensity, velocity centroid, and velocity channel maps. These maps, derived from 3D magnetohydrodynamic (MHD) turbulence simulations, serve as inputs for the CNN training. By learning the relationship between these structural features and the underlying turbulence properties, CNN can predict $M_s$ under various conditions, including different magnetic fields and levels of observational noise. The median uncertainty of the CNN-predicted $M_s$ ranges from 0.5 to 1.5 depending on the noise level. While intensity maps offer lower uncertainty, channel maps have the advantage of predicting the 3D $M_s$ distribution, which is crucial in estimating 3D magnetic field strength. Our results demonstrate that machine-learning-based tools can effectively characterize complex turbulence properties in the ISM.
\end{abstract}



\section{Introduction}
MHD turbulence plays a pivotal role as an influential factor shaping the evolution and structure of the interstellar medium (ISM; \citealt{1981MNRAS.194..809L,1995ApJ...443..209A, 2010ApJ...710..853C,2017ApJ...835....2X,2022ApJ...934....7H,2019NatAs...3..776H,2022ApJ...941..133H}). Namely, it contributes to many astrophysical processes in the ISM including but not limited to molecular cloud evolution \citep{2001ApJ...546..980O,2008ApJ...688L..79F,2012ApJ...757..154L,2014SSRv..181....1L,2022MNRAS.513.2100H,2024MNRAS.530.3431V}, star formation \citep{MK04,Crutcher04,annurev:/content/journals/10.1146/annurev.astro.45.051806.110602,2012ApJ...761..156F,Hu_2021,Hu_2022}, 
cosmic ray propagation \citep{1966ApJ...146..480J,2011ApJ...741...16G,2013ApJ...779..140X, 2022MNRAS.512.2111H,2022ApJ...934..136X}, 
characterization of the B-mode polarizations in the
Cosmic Microwave Background (CMB; \citealt{Pogosian_2014,Bracco_2019}). Despite MHD turbulence's critical importance, our understanding of astrophysical implications remains incomplete.

Particularly for star formation, MHD turbulence can provide global support against the forces of gravity in molecular clouds \citep{1993ApJ...419L..29E,10.3389/fspas.2018.00039,Mignon-Risse2023,2024MNRAS.530.3431V} but produce local density fluctuations serving as seeds of star formation  \citep{MK04,2005ApJ...630..250K,annurev:/content/journals/10.1146/annurev.astro.45.051806.110602,2018ApJ...863..118B}.
The sonic Mach number, $M_s$, is a critical parameter in characterizing MHD turbulence, representing the ratio of the turbulent velocity to the sound speed. This ratio is indicative of fluid compressibility and is crucial for understanding star formation in molecular clouds. Over the years, a number of methods have been developed to measure $M_s$ alongside other turbulent parameters like the Alfv\'enic Mach number, $M_A$. One common approach involves analyzing the column density probability distribution functions (N-PDFs), which generally assume a lognormal distribution in regions dominated by turbulence \citep{1994ApJ...423..681V, Hill_2008, 10.1111/j.1745-3933.2008.00526.x, Brunt2010, 2011MNRAS.416.1436B, 2015MNRAS.448.3297F, 2018ApJ...863..118B}. The width or dispersion of the N-PDFs is closely related to $M_s$; a larger 
$M_s$ represents significant density fluctuations and consequently, a broader N-PDF. Several studies have explored the potential of estimating $M_s$ from the skewness of the N-PDF \citep{Kowal_2007, Burkhart_2010} and from the variance of column density \citep{1997MNRAS.288..145P, PhysRevE.58.4501, Burkhart_2012}. Additionally, some studies have applied more complex functions, like the Tsallis function, to fit the N-PDFs. By analyzing the parameters of these fits, further estimations of $M_s$ can be achieved \citep{Esquivel_2010, Tofflemire_2011}. These statistical measures have provided crucial insights into the MHD turbulence's properties within the cloud.

However, determining the spatial distribution of the $M_s$ on the plane-of-the-sky (POS) still poses significant challenges, particularly due to the need for large data samples for the typical statistical methods. Estimating $M_s$ in multi-phase ISM, such as neutral hydrogen (H I) and H$\alpha$ regions, adds further complexity. Additionally, variations in $M_s$ along the line of sight (LOS) are critical for assessing the 3D distribution of magnetic field strength in these media \citep{Lazarian_2022,Hu_2023,2024ApJ...974..237L}. In this work, we introduce a machine learning paradigm for estimating the $M_s$, since the morphological changes of density and intensity structures are associated with different $M_s$ values. Highly supersonic media, characterized by significant small-scale density fluctuations, typically exhibit more pronounced small-scale filamentary structures. These morphological variations in integrated intensity maps of spectroscopic observations have been previously noted \citep{2019ApJ...878..157X,2019ApJ...886...17H,10.1093/mnras/stz3377,2021ApJ...910...88X,2020ApJ...905..129H}. Earlier work by \cite{2019ApJ...882L..12P} 
and \cite{hu2023probing} has demonstrated that Convolutional Neural Networks (CNNs; \citealt{726791}) are effective in extracting morphological features from spectroscopic observations and correlating these features with the $M_A$. Our current research builds on this foundation, marking the first attempt to use CNNs combined with spectroscopic observations to estimate $M_s$. 
 
Moreover, the CNN approach extends beyond analyzing integrated intensity or column density maps. It also includes the examination of velocity centroid maps and velocity channels \citep{Lazarian_2000,10.1093/mnras/stw1296,2021ApJ...910..161Y,2023MNRAS.524.2994H}, which offer additional insights into $M_s$ based on velocity information \citep{2020ApJ...898...65Y}. Applying CNNs to H I observations, particularly when combined with the velocity and spatial information provided by the Galactic rotation curve, enables modeling the 3D distribution of $M_s$ across the Milky Way \citep{Hu_2023}. In this work, we train our CNN with synthetic spectroscopic observations generated from 3D simulations of MHD turbulence. We will compare the effectiveness of using intensity, centroid, and channel maps to estimate $M_s$.

This work is organized into five sections. \S~\ref{sec:nummeth} describes the numerical MHD simulations, synthetic spectroscopic observations, and training strategies of the CNN. \S~\ref{sec:results} details the results of the CNN approach at predicting $M_s$ numbers, including testing the effects of noise, magnetic field inclination, and missing low-spatial frequencies. 
\S~\ref{sec:discussion} discusses our results and their implications, and finally \S~\ref{sec:conclusion} concludes our paper and provides future outlooks of this research.

\section{Numerical Methods}
\label{sec:nummeth}
\subsection{MHD simulations}
This study utilized MHD simulations which were generated from ZEUS-MP/3D \citep{Hayes_2006}. We solve the ideal MHD equations in an Eulerian framework with associated periodic boundary and isothermal conditions. The simulation is regularly gridded into $512^3$ cells.

The scale-free simulation of MHD turbulence can be characterized via the sonic Mach number $M_s=\delta v_{\rm inj}/c_s$ and Alfv\'enic Mach number $M_A=\delta v_{\rm inj}/v_A$, where $v_{\rm inj}$ is the turbulent velocity at injection scale, $c_s$ is the sound speed, and $v_A$ is the Alfven speed. Initially, a uniform density field and magnetic field are set up and the magnetic field orientates perpendicular to the LOS. For this work, simulations were chosen based on having similar $M_A$ values but different $M_s$ values to feed the CNN with different turbulence conditions. 

Turbulence is solenoidally driven at wavenumber $k=2$ to achieve a Kolmogorov spectrum. We change the injection velocity $v_{\rm inj}$ to produce a wide variety of $M_s$ values, as listed in Tab.~\ref{tab.sim}. Similar MHD simulations have been used in \cite{2024arXiv240407806H}. We refer readers to \cite{2024arXiv240407806H} for more details on the simulation.

\begin{table}
\begin{center}
\centering
\begin{tabular}{|c|c|c|c|}
 \hline
 Simulation & $M_s$ & $M_A$ & Range of $M_s^{\rm sub}$ \\
 \hline\hline
 min01 & 2.17 & 0.51 & 1.36 - 2.84 \\ 
 \hline
 min02 & 4.35 & 0.52 & 2.78 - 6.32 \\
 \hline
 min03 & 6.53 & 0.52 & 3.85 - 8.97  \\
 \hline
 min04 & 8.55 & 0.51 & 5.07 - 11.46 \\
 \hline
 min05 & 10.86 & 0.69 & 6.18 - 13.68 \\
 \hline
 min06 & 11.02 & 0.43 & 7.11 - 15.60 \\
 \hline
\end{tabular}
\caption{$M_s$ and $M_A$ are the sonic Mach number and the Alfv\'enic Mach number calculated from the global injection velocity, respectively. $M_{s}^{\rm sub}$ values are determined using the local velocity dispersion calculated along each LOS in a $32\times32$ cells sub-field. }
\label{tab.sim}
\end{center}
\end{table}

\begin{figure*}
	\includegraphics[width=1.0\linewidth]{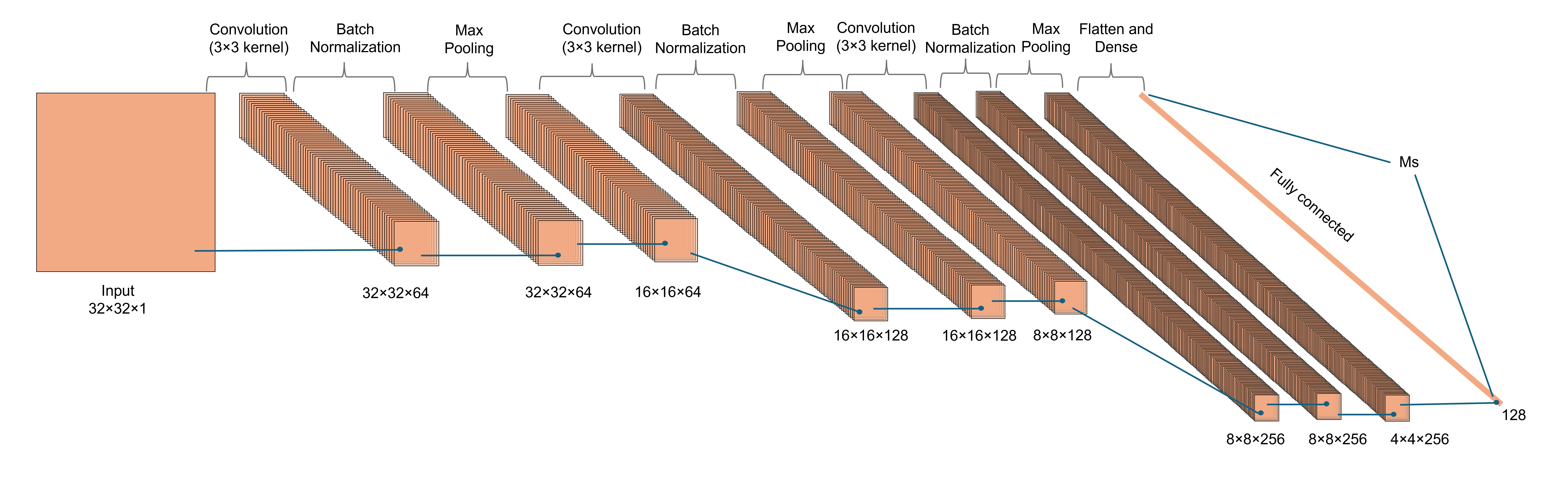}
    \caption{CNN model architecture utilized in this study. The input image is a 32 $\times$ 32 cells map cropped from either a channel, centroid, or intensity map. The CNN model then outputs a prediction of the sonic Mach numbers $M_s$.}
    \label{fig:Convolution Neural Netowrk}
\end{figure*}

\subsection{Intensity, centroid, and channel maps}
To create synthetic spectroscopic observations conducive to our CNN training, we utilize velocity and density fields derived from MHD simulations to construct Position-Position-Velocity (PPV) cubes. From these PPV cubes, we generate intensity maps, velocity centroids, and velocity channel maps, each serving distinct purposes in our model training process.

Intensity maps are produced by fully integrating the PPV cubes along the LOS. This integration eliminates all velocity information, yielding a map that reflects only the density or intensity distribution. The intensity map, $I(x,y)$, is defined mathematically as:
\begin{equation}
I(x, y) = \int \rho(x, y, v_{\rm los}) dv_{\rm los},
\end{equation}
where $\rho(x,y,v_{\rm los})$ represents the intensity values within the PPV cube, and 
$v_{\rm los}$ is the LOS velocity. $\rho(x,y,v_{\rm los})$ corresponds to brightness or antenna temperature in observation.
  
Velocity centroids, $C(x,y)$, provide an intensity-weighted average velocity for each cell in the map, integrated along the LOS. They are calculated from:
\begin{equation}
C(x, y) = \frac{\int \rho(x, y, v_{\rm los}) v_{\rm los} dv_{\rm los}}{\int \rho(x, y, v_{\rm los}) dv_{\rm los}}.
\end{equation}

Velocity channel maps, $Ch(x,y)$, are derived by integrating the PPV cube over a narrowly defined velocity range. Unlike intensity maps, these channel maps are predominantly affected by velocity fluctuations due to the phenomenon known as velocity caustics \citep{Lazarian_2000,10.1093/mnras/stw1296,2021ApJ...910..161Y,2023MNRAS.524.2994H}. This effect arises when gas at different LOS positions but with similar LOS velocities is sampled into the same location in PPV space, leading to crowded and morphologically distorted intensity distributions. 

The expression for a thin velocity channel is given by:
\begin{equation}
    Ch(x,y)= \int^{v_0+\Delta v/2}_{v_0-\Delta v/2} \rho(x,y,v_{\rm los})dv_{\rm los},
\end{equation}
where 
$v_0$ is the center velocity of the channel and $\Delta v$ is the channel width, which is chosen to be less than the square root of the turbulence velocity dispersion, $\Delta v<\sqrt{(\delta v)^2}$.

These three maps have different weights of density and velocity information. We explore and compare them in terms of estimating $M_s$ with the CNN model.

\subsection{Convolutional Neural Network (CNN)}
In this work, we adopt a CNN model to estimate $M_s$. The CNN model has been successfully applied to spectroscopic observation and radio observation to estimate $M_A$ \citep{hu2023probing,2024arXiv240407806H,2024arXiv241009294H,2024arXiv241107080Z}. The CNN architecture consists of initial layers that stack on top of each other and make up convolution layers. This is then followed by pooling layers, in our case max pooling. Batch normalization layers follow each convolution layer and these layers help induce faster convergence during the training process. A prediction can be extracted and processed by the fully connected layers and outputs a prediction of the sonic Mach number.  This CNN model is given in Fig.~\ref{fig:Convolution Neural Netowrk}.

The target for training was the $M_s$. We use the Mean-Squared Error (MSE) loss function for backpropagation. For each training iteration, approximately 0.6 million sub-fields were used. We conducted at least 20 training iterations, continuing until the MSE loss function indicated saturation, suggesting that the model parameters had converged effectively.

\subsection{Synthetic $M_s$ maps}
Our training input is either the intensity map, velocity centroid map, or velocity channel map generated from the simulations. The input maps are normalized by their maximum values so only morphological features in the map are the most important. For each $32\times32$-cells sub-field, we also generate corresponding $M_s^{\rm sub}$ maps as per the following:
\begin{equation}
\begin{aligned}
M_s^{\rm sub}&=\frac{v^{\rm sub}}{c_s},
\end{aligned}
\end{equation}
where $M_s^{\rm sub}$ is defined using the local velocity dispersion for each sub-field (i.e., $v^{\rm sub}$), rather than the global turbulent injection velocity $v_{\rm inj}$ used to characterize the full simulation. The ranges of $M_s^{\rm sub}$ averaged over the sub-field in each simulation are listed in Tab.~\ref{tab.sim}. These values of $M_s^{\rm sub}$ cover typical physical conditions of the ISM.

\subsection{Network training}
The CNN's trainable parameters undergo optimization by following a typical neural network training approach. 

{\bf Random cropping:}
While training the CNN, we apply a strategy that helps with diversifying the training dataset which improves the generalization ability of the CNN. This strategy works by altering input images or data by applying random cropping to them which results in cell sizes of 32 $\times$ 32 \citep{Takahashi_2020}. This random cropping method incorporates variability and randomness into the training data \citep{hu2023probing,2024arXiv240407806H,2024arXiv241009294H}.

{\bf Random rotation:}
Another strategy employed in this work is the use of random rotations. Images that are utilized to train the model lack rotational invariance. In the CNN's view, each cell is directly related to an element in a matrix. By rotating the image, the matrix's elements are rearranged and create a new image for the model to train on \citep{Larochelle2007}. This fact can be utilized in two ways. The first is that by randomly rotating our 32 $\times$ 32-cells, the training dataset can be further augmented. The second is that the original non-rotated datasets can be used as validation and a method of creating prediction test scenarios. In effect, random rotation enhances the training data with more diversity and randomness to improve the accuracy of the CNN's predictive capabilities across a wide variety of physical situations in the ISM \citep{vanDyk2001}.

\begin{figure*}
  \centering
	\includegraphics[width=1.0\linewidth]{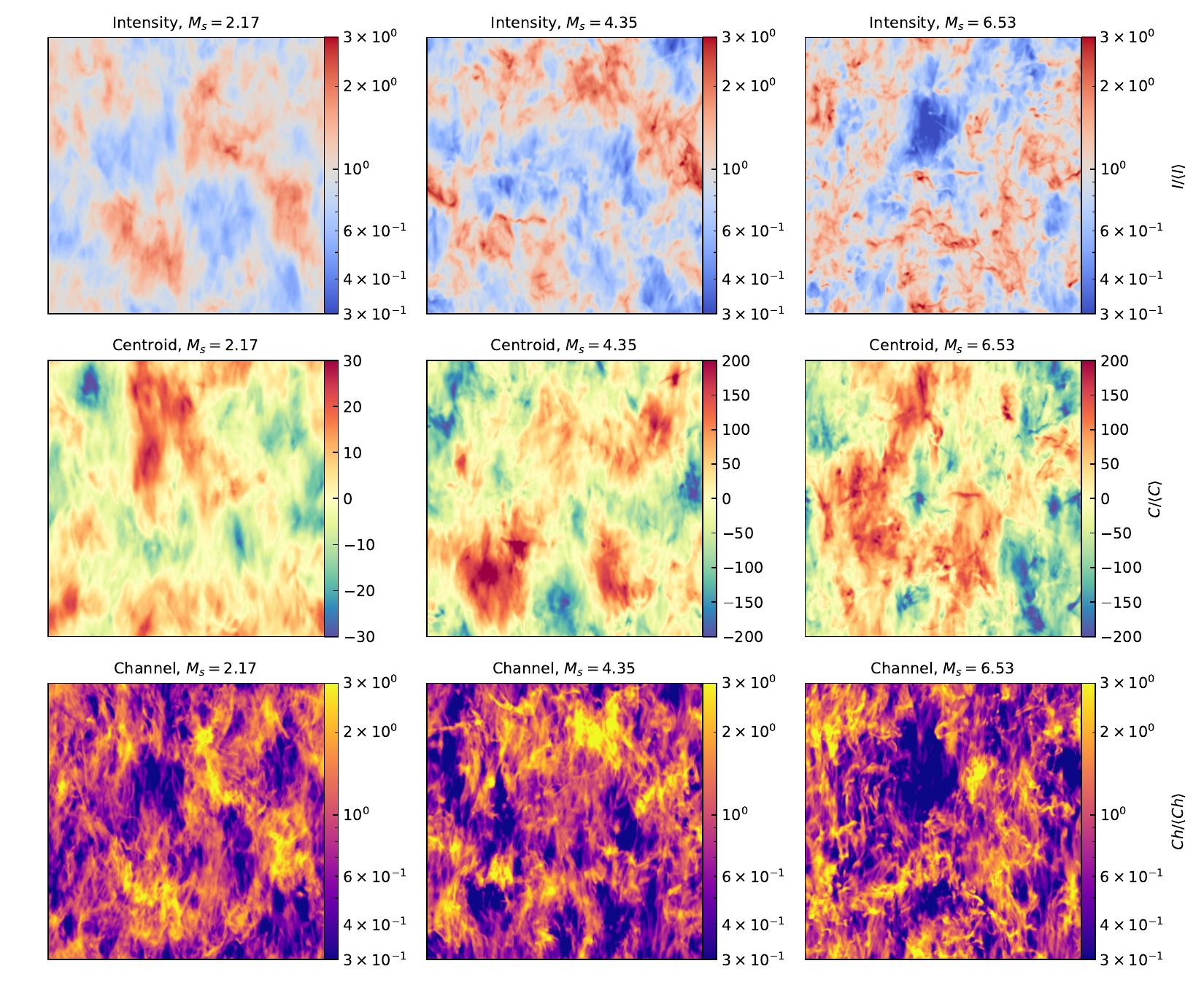}
    \caption{Maps of the integrated intensity (top), velocity centroid (middle), and thin velocity channel (bottom). Three simulations with different $M_s=2.17,4.35,$ and 6.53 are given here for comparison.  From left to right $M_s$ is increasing. }
    \label{fig:Intensity Channel Centroid Maps Figure}
\end{figure*}

\begin{figure*}
 \centering
	\includegraphics[width=.7\linewidth]{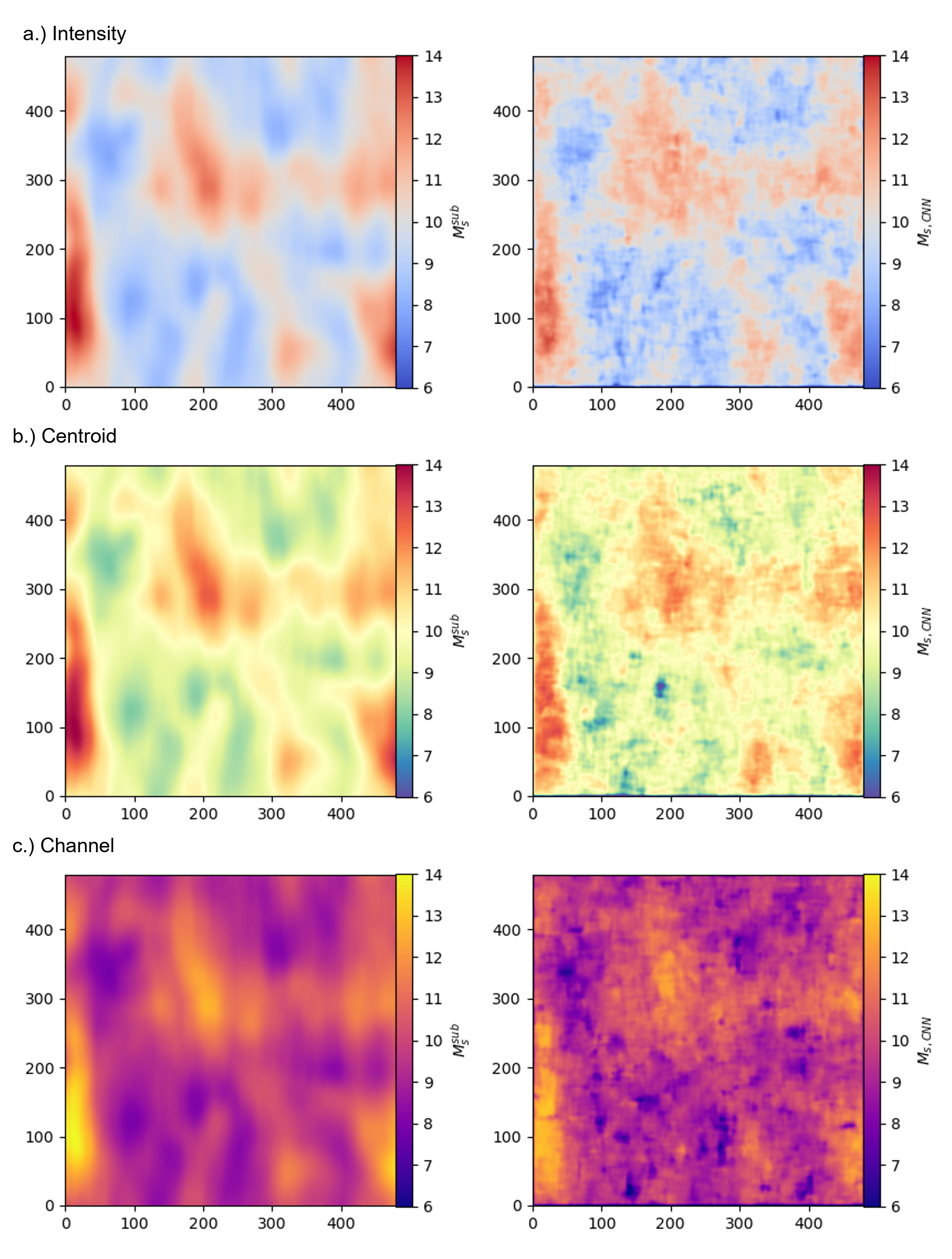}
    \caption{The right column presents a comparison of the CNN-predicted sonic Mach number $M_{s, {\rm CNN}}$ distribution from intensity (top), centroid (middle), and channel input maps (bottom) and the actual $M_{s}^{\rm sub}$ 
    distribution in the left column. The supersonic simulation with  $M_s=11.02$ is used as an example here.}
    \label{fig:ms90 min02 and input maps 90 degrees prediction min02}
\end{figure*}

\begin{figure*}
	\includegraphics[width=1.0\linewidth]{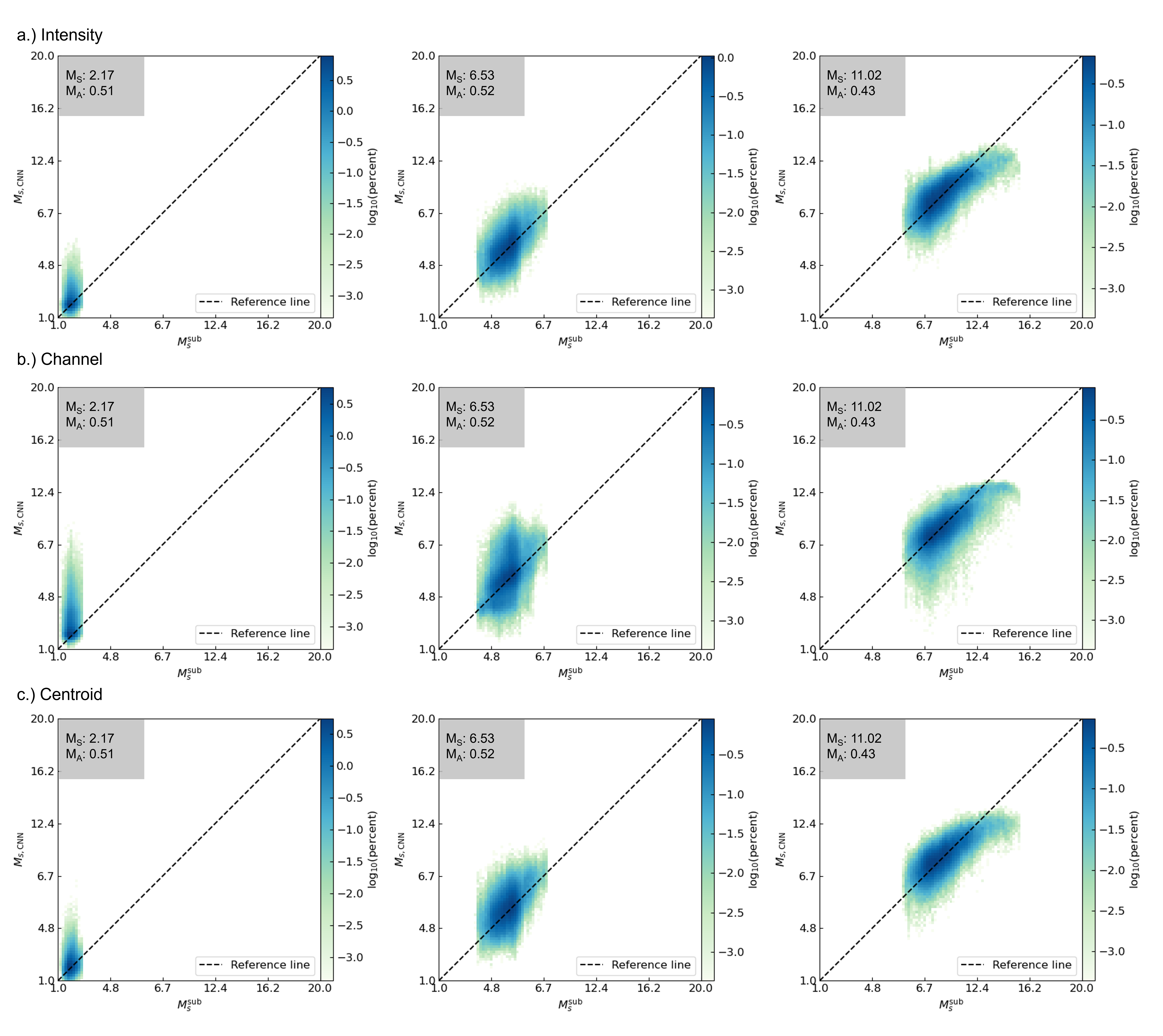}
    \caption{2D histograms comparing CNN-predicted $M_{s, {\rm CNN}}$ and actual $M_s$ of three simulations $M_s=2.17,6.53$, and 11.02. $M_{s, {\rm CNN}}$ is predicted from either intensity map (top), velocity channel map (middle), or centroid map (bottom). The dashed reference line refers to the one-to-one correlation.}
    \label{fig:90 Degrees Histograms}
\end{figure*}
\section{Results}
\label{sec:results}
\subsection{Intensity, centroid, and channel maps with different $M_s$}
Fig.~\ref{fig:Intensity Channel Centroid Maps Figure} presents three examples of intensity, velocity centroid, and velocity channel maps for different sonic Mach numbers, $M_s = 2.17$, 4.35, and 6.53, while maintaining similar Alfv\'enic Mach numbers, $M_A$. The intensity maps clearly show that intensity structures undergo significant changes as $M_s$ increases. At a low $M_s$ value of 2.17, intensity fluctuations are primarily concentrated on large scales due to the relatively weak influence of shocks. Their statistics are passively regulated by those of velocity fluctuations \citep{2005ApJ...624L..93B}. The resulting intensity structures tend to form coherent, large-scale filamentary patterns. However, as $M_s$ increases, the impact of shocks becomes more pronounced, generating more small-scale, high-intensity fluctuations. This is particularly evident in the intensity map for $M_s = 6.53$, where the intensity structure shifts towards smaller-scale filamentary features. One could expect a large aspect ratio (width to height) of those structures in a high $M_s$ medium. These results suggest that variations in $M_s$ are distinctly imprinted on the intensity maps.

In the case of centroid maps, similar trends with varying $M_s$ are observed. Compared to intensity maps, centroid maps incorporate intensity-weighted velocity information. At low $M_s$, the statistical behavior of velocity fluctuations closely resembles that of density fluctuations \citep{2005ApJ...624L..93B}, resulting in centroid structures that are morphologically similar to intensity structures. As $M_s$ increases, small-scale fluctuations also become more prominent, though to a lesser extent than in the intensity maps. This is expected, as shocks tend to dissipate small-scale velocity fluctuations, leading to a steep velocity spectrum with a slope of approximately -2. Although centroid maps primarily reflect velocity information, the intensity weighting introduces some dependency on intensity, causing the structures to be influenced by changes in $M_s$.

In contrast, structures in channel maps are much less sensitive to variations in $M_s$. Regardless of the increase in $M_s$, channel maps consistently display highly filamentary small-scale structures. This behavior is primarily due to the dominance of velocity caustics in thin velocity channels \citep{Lazarian_2000,10.1093/mnras/stw1296,2021ApJ...910..161Y,2023MNRAS.524.2994H}. The statistics of these maps are therefore primarily governed by velocity fluctuations, though density fluctuations also contribute. Compared to intensity and centroid maps, thin channel maps exhibit less sensitivity to changes in $M_s$. 

\begin{figure*}
    \centering
    \includegraphics[width=1.0\linewidth]{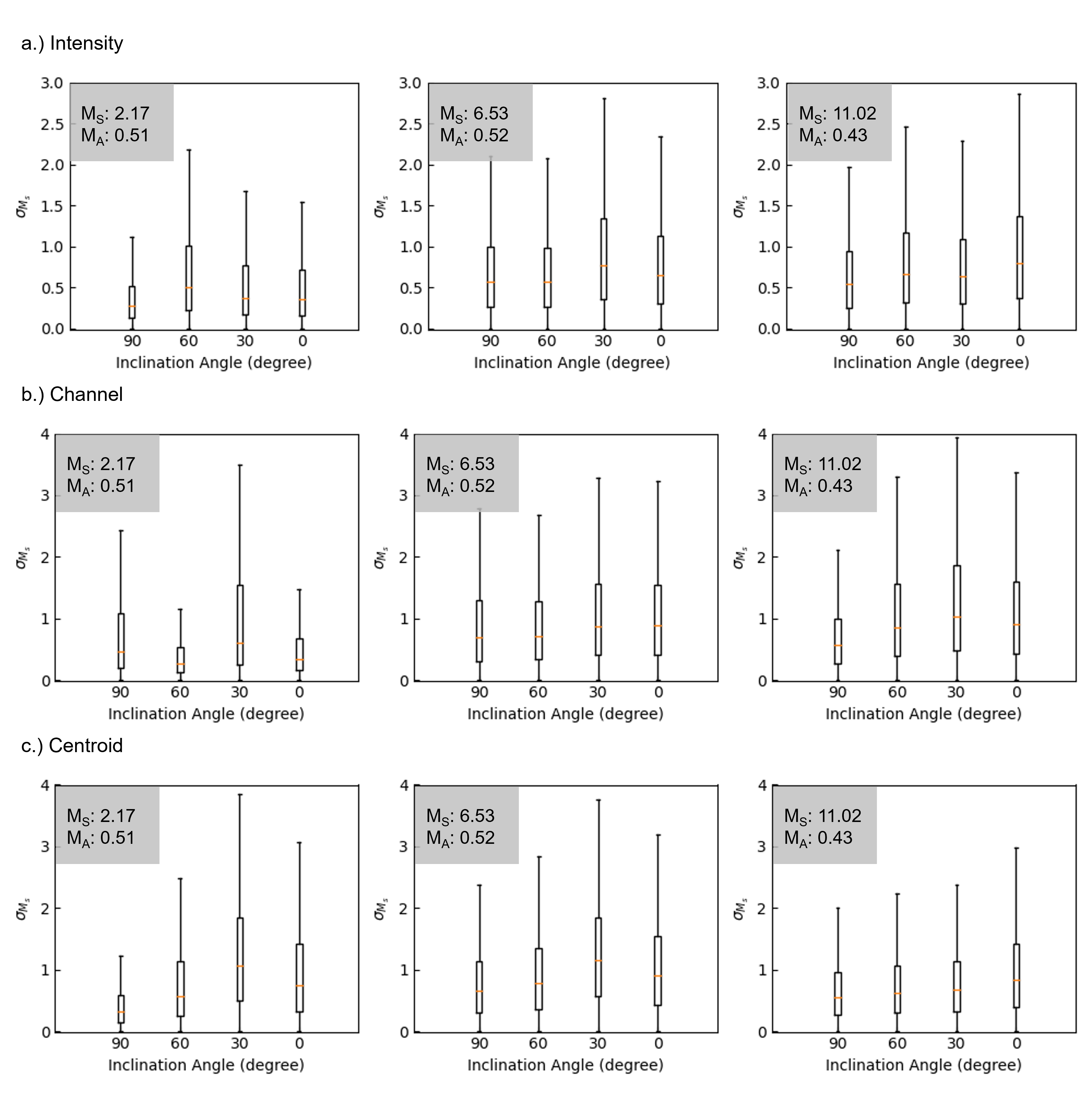}
    \caption{Box plots of the absolute difference $\sigma_{M_s}$ between CNN-predicted $M_s$ and actual $M_{s}^{\rm sub}$ values for various inclination angles of mean magnetic field. Three simulations $M_s=2.17$, 6.53, and 11.02 are used. $M_{s, {\rm CNN}}$ is predicted from either intensity map (top), velocity channel map (middle), or centroid map (bottom). The box provides the range of the first quartile (lower) to the third quartile (upper), the upper and and lower line gives the maximum and minimum error respectively, and the orange line represents the median value.}
    \label{fig:Boxplots Figure}
\end{figure*}

\begin{figure*}
    \centering
    \includegraphics[width=1.0\linewidth]{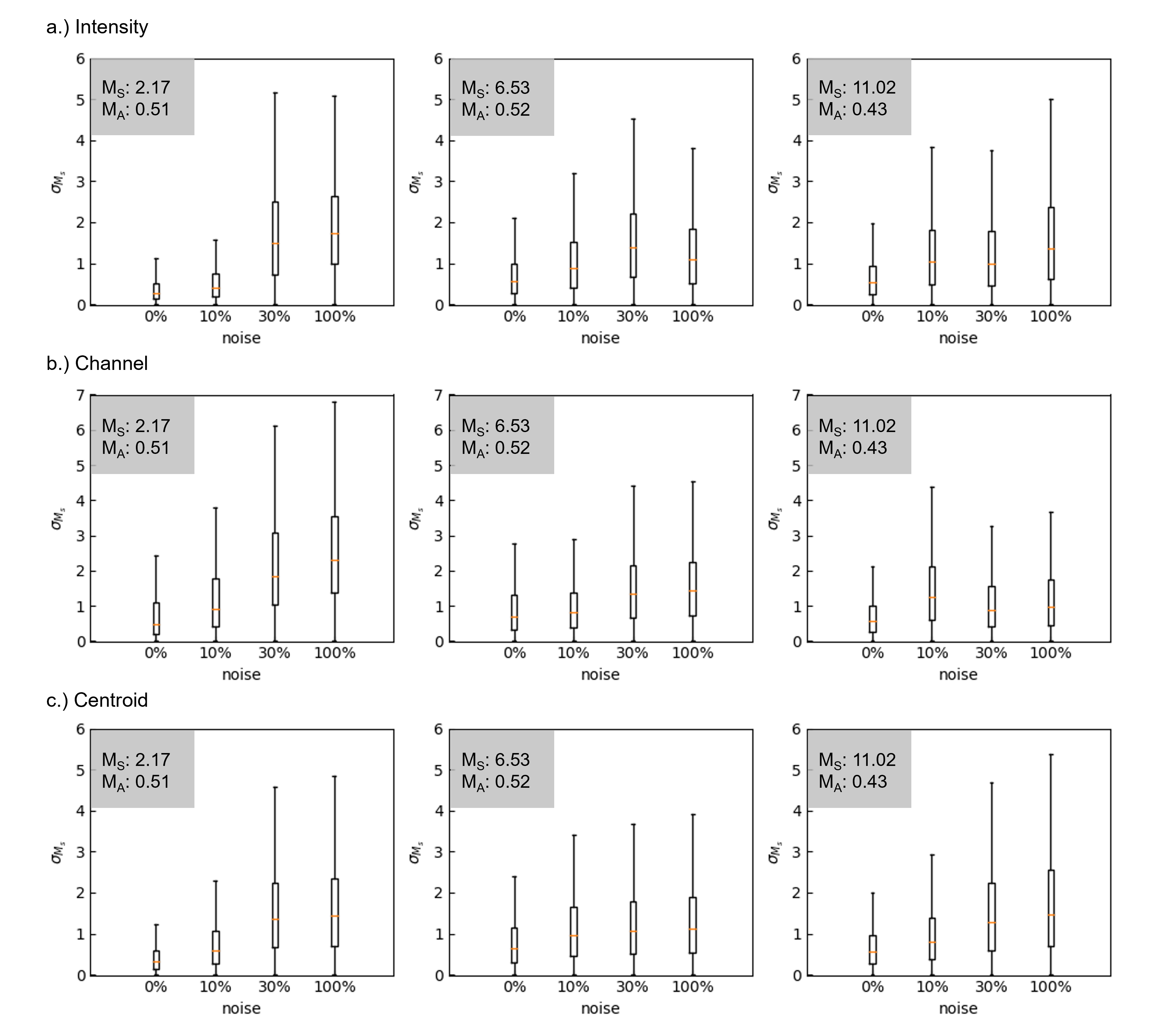}
    \caption{Box plots of the absolute difference $\sigma_{M_s}$ between CNN-predicted $M_s$ and actual $M_{s}^{\rm sub}$ values for various levels of noise. Three simulations $M_s=2.17$, 6.53, and 11.02 with an inclination angle of 90 degrees are used. $M_{s, {\rm CNN}}$ is predicted from either intensity map (top), velocity channel map (middle), or centroid map (bottom). The box provides the range of the first quartile (lower) to the third quartile (upper), the upper and lower line gives the maximum and minimum error respectively, and the orange line represents the median value.}
    \label{fig:Noise Boxplots Figure}
\end{figure*}
\begin{figure*}
    \centering
    \includegraphics[width=1.0\linewidth]{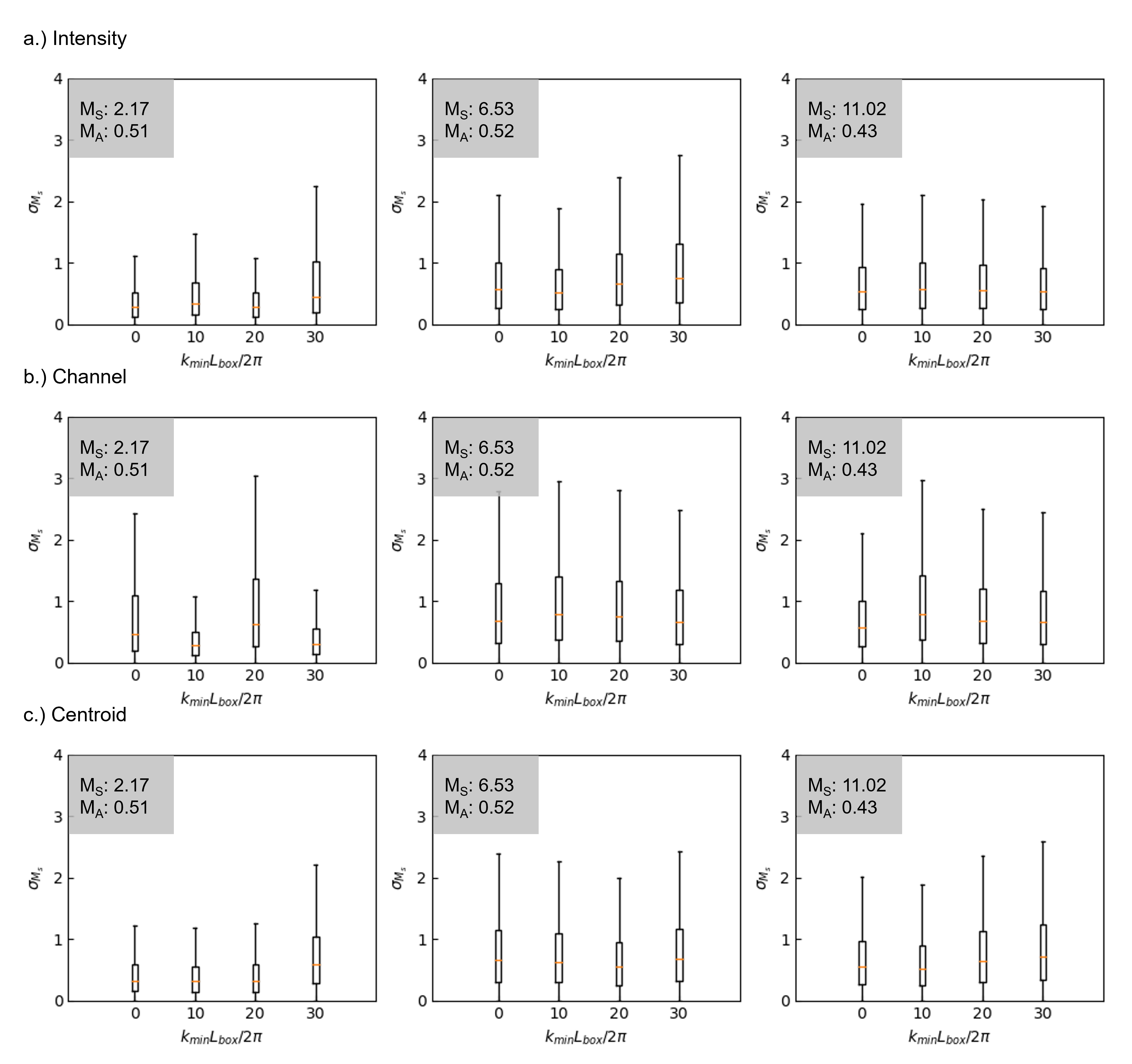}
    \caption{Box plots of the absolute difference $\sigma_{M_s}$ between CNN-predicted $M_s$ and actual $M_{s}^{\rm sub}$ values for various levels of removed spatial frequencies. $K_{\rm min}$ represents minimum wavenumber remaining in the filtered input map and $L_{\rm box}$ is the size of simulation box. Each of the three simulations $M_s=2.17$, 6.53, and 11.02 were studied for the 90 degree inclination angle case. $M_{s, {\rm CNN}}$ is predicted from either intensity map (top), velocity channel map (middle), or centroid map (bottom). The box provides a range of errors from the first quartile (lower) to the third quartile (upper). The upper and lower line gives the maximum and minimum error respectively, and the orange line represents the median $\sigma_{M_s}$ error value.}
    \label{fig:Spatial Frequency Boxplots Figure}
\end{figure*}

\subsection{Sonic Mach Number Predictions}
Fig.~\ref{fig:ms90 min02 and input maps 90 degrees prediction min02} presents a comparison between the CNN-predicted sonic Mach number distribution, $M_{s, {\rm CNN}}$, and the actual distribution, $M_{s}^{\rm sub}$. The input to the CNN model is either an intensity map, centroid map, or channel map generated from the simulation $M_s = 11.02$. Overall, for all three kinds of input maps, their predicted $M_{s, {\rm CNN}}$ distribution closely matches $M_{s}^{\rm sub}$, although local discrepancies are present \footnote{ Note that the uncertainty in CNN predictions increases if the data are unseen or if the conditions are not included in the training process (see Appendix~\ref{appendix2}). An unsupervised machine learning model might be used to improve the accuracy of unseen data.}. Another example of predicting the $M_{s, {\rm CNN}}$ in trans-/super-Alfv\'enic condition with $M_A$ = 1.07 is given in Appendix~\ref{appendix}. The results are similar to those from the sub-Alfv\'enic condition, suggesting the CNN approach works for both strongly and weakly magnetized media.

Fig.~\ref{fig:90 Degrees Histograms} displays 2D histograms of $M_{s, {\rm CNN}}$ and $M_{s}^{\rm sub}$. Generally, for three kinds of input maps (intensity, centroid, and channel maps) the proximity of the data points to the one-to-one reference line indicates a statistical agreement between predicted and true values, highlighting the CNN model's effectiveness, albeit with some scatter that reflects deviations from the actual values. The 2D histograms show that the CNN model overestimates for the low $M_s=2.17$ case, while it is more likely to underestimate for the high $M_s=11.02$ case. 

\subsection{Effect of the mean magnetic field's inclination}
The CNN approach for predicting the $M_s$ distribution is based on the morphological changes in intensity, centroid, and channel structures under varying $M_s$ conditions. Another factor that may influence the observed structures is the inclination angle of the mean magnetic field. In a magnetized turbulent medium, density and velocity structures tend to elongate along magnetic fields in subsonic conditions, whereas, in the supersonic regime, small-scale density structures can become perpendicular to the magnetic fields due to shock effects \citep{2019ApJ...886...17H,2019ApJ...878..157X}. This elongation is a well-known consequence of the anisotropy of MHD turbulence \citep{1995ApJ...438..763G,1999ApJ...517..700L}. When the mean magnetic field is perpendicular to the LOS, this elongation is most pronounced, as shown in Fig.~\ref{fig:Intensity Channel Centroid Maps Figure}, where the mean magnetic field is aligned along the vertical axis. However, as the inclination angle between the mean magnetic field and the LOS decreases, the observed structures change. In particular, when the mean field is fully parallel to the LOS, the elongation effect cannot be observed.

To investigate the effect of the mean field's inclination, we rotate the simulation boxes to achieve four different inclination angles (90, 60, 30, and 0 degrees), generate synthetic spectroscopic data, and retrain the CNN model accordingly. We quantify the CNN prediction error by calculating the absolute difference between the predicted and actual $M_s$ values, denoted as $\sigma_{M_s}$. Fig.~\ref{fig:Boxplots Figure} presents box plots of $\sigma_{M_s}$ for three simulations with $M_s = 2.17$, 6.53, and 11.02, across the four inclination angles (90, 60, 30, and 0 degrees). This box plot analysis was conducted for all three types of input maps—intensity, centroid, and channel maps—used to make the predictions.

For predictions based on all three map types shown in Fig.~\ref{fig:Boxplots Figure}, the median error is generally around 0.5 when the inclination angle is 90 degrees, while for the $M_s = 2.17$ case, the median error is lower, ranging from 0.2 to 0.3. Although maximum or upper quartile error values vary more at higher $M_s$, the median error remains approximately $\sigma_{M_s} = 0.5$ across simulations with different $M_s$ values. When comparing errors calculated from intensity, centroid, and channel maps, while their median errors are similar, the maximum and upper quartile error values tend to be lowest for intensity maps and highest for channel maps, especially for the lower $M_s = 2.17$ and $M_s = 6.53$ cases.

For the 60, 30, and 0 degrees inclination angles, where the magnetic field is not perpendicular to the LOS, the error exhibits more variations, with effectively higher maximum and upper quartile values. As seen in Fig.~\ref{fig:Boxplots Figure}, as the inclination angle decreases from 90 to 0 degrees, the error $\sigma_{M_s}$ generally increases. This is particularly evident for predictions from channel maps, where the error is highest for the 0 degrees and 30 degrees cases. In contrast, predictions from intensity maps appear less sensitive to the inclination angle. Nevertheless, the median errors remain generally below 0.75 across all inclination angles, highlighting the robustness of our CNN architecture in accurately predicting $M_s$ values.

\subsection{Noise effect}
Noise is unavoidable in observation and may diminish useful information in the intensity, centroid, and channel maps. In this section, we examine the impact of noise on our CNN model's performance. We introduce Gaussian noise to the maps, with the noise level quantified by the noise ratio (NR), defined as the ratio of the Gaussian noise's standard deviation to the mean value of the intensity, channel, or centroid maps.

Fig.~\ref{fig:Boxplots Figure} presents the absolute error $\sigma_{M_s}$ for three simulations with $M_s = 2.17$, 6.53, and 11.02 across four different NR values: 0\%, 10\%, 30\%, and 100\%. This analysis is conducted for three types of input maps—intensity, centroid, and channel maps, all with an inclination angle of 90 degrees—on which the CNN model is trained to make predictions. In the noise-free case, the median error remains low, with the 0\% NR median $\sigma_{M_s}$ values staying below 0.75 for all input maps. The results for the 0\% NR case, which corresponds to the 90-degree case for each input map, were discussed in the previous section. Across all input map types and $M_s$ values, the median error remains generally below 2, despite increasing with rising noise levels.

In simulations with added noise (NR values of 10\%, 30\%, and 100\%), the error increases relative to the noise-free case. Across all input map types and $M_s$ values, the median $\sigma_{M_s}$ values rise as the noise ratio increases. In addition to the median errors, the upper quartile values also increase, with the largest errors occurring in the 30\% and 100\% NR simulations for all map types. For intensity maps, the highest median error occurs for the $M_s=2.17$ case at 100\% NR, though similarly high values are observed for the $M_s=6.53$ case at 30\% NR and the $M_s=11.02$ case at 100\% NR. Notably, for intensity maps, the median $\sigma_{M_s}$ never exceeds 2 in any of the noise-added simulations.

For channel maps, $M_s=2.17$ shows the highest median and upper quartile $\sigma_{M_s}$ values, with $\sigma_{M_s}$ exceeding 2 in the 100\% NR case and approaching 2 in the 30\% NR case. The $M_s=11.02$ case for channel maps displays the highest median error for the 10\% NR model across all input map types and $M_s$ values. However, for both $M_s=6.53$ and $M_s=11.02$, the median $\sigma_{M_s}$ values generally remain below 2 for all map types. For centroid maps, the trend of increasing median $\sigma_{M_s}$ with rising noise levels persists. Although the 100\% NR models for all three $M_s$ values show the highest error values, these errors remain below 2.

The effectiveness of our CNN model in accurately predicting $M_s$ values is evidenced by the consistently low median $\sigma_{M_s}$ across all input map types and noise conditions. As expected, increased noise levels generally led to higher median and upper-quartile error values. Specifically, the median errors rose from less than 0.75 in the noise-free cases to approximately 2 – 2.25 under the highest noise conditions. These increases make predictions more challenging for lower $M_s$ values, such as in our simulation with $M_s = 2.17$. However, for higher $M_s$ values, such as $M_s > 10$, a median error of around 2, even in the presence of 100\% noise, remains relatively favorable.

\subsection{Removal of Lower Spatial Frequencies}
For extra-galactic spectroscopic observations, telescope arrays like the Very Large Array (VLA) are often used. 
Due to limited baselines, these observations are usually constrained by the loss of low spatial frequencies which are necessary to resolve large-scale features of structures within images. In this study, we tested our CNN model's predictive ability when lower spatial frequencies are removed. We applied a $k$-space filter to the intensity, centroid, and channel maps before CNN training, following these steps: (i) performing a Fast Fourier Transform (FFT) on the 2D map, (ii) filtering out intensity values from wavenumber $k=0$ to $k_{\rm min}$ to emphasize high-spatial frequencies, and (iii) applying an inverse FFT to transform the filtered map back into the spatial domain. We explored four different ranges of removed spatial frequencies—0, 0 - 10, 0 - 20, and 0 - 30—across three different $M_s$ simulations for each of the three input map types.

Fig.~\ref{fig:Spatial Frequency Boxplots Figure} presents box plots of the prediction error corresponding to the different input maps, $M_s$ simulations, and removed spatial frequency ranges used for CNN model training. Across all input map types, the median error is less than one, as shown in Fig.~\ref{fig:Spatial Frequency Boxplots Figure}. There is a general trend of increasing median and upper quartile error values as more spatial frequencies are removed, particularly for the intensity and centroid input maps. However, the channel input maps exhibit some irregularities in this trend.

For the intensity input maps, the $M_s = 2.17$ simulation shows the lowest overall median error, with values below 0.5. As more spatial frequencies are removed, the prediction error increases, reaching its highest maximum error for the 0 - 30 spatial frequency removal case. Despite this, the median $\sigma_{M_s}$ remains below one. As higher $M_s$ simulations are used for training, there is an increase in median $\sigma_{M_s}$ values. The $M_s = 11.02$ simulation shows an almost constant median error across all spatial frequency regimes, with values consistently below one.

In the centroid input map case, the median $\sigma_{M_s}$ remains stably below one, with a general trend of increasing $\sigma_{M_s}$ as the simulation regime transitions from low $M_s$ to high $M_s$. The $M_s = 2.17$ simulation shows the lowest errors, with the 0 - 10 and 0 - 20 removed spatial frequency cases producing $\sigma_{M_s}$ values below 0.5. Both $M_s = 6.53$ and $M_s = 11.02$ display similar median $\sigma_{M_s}$ values, ranging between 0.5 and one across all removed spatial frequency instances.

In the channel input map predictions, more erratic behavior is observed. Unlike the intensity and centroid maps, no consistent trend emerges across the three $M_s$ simulations, with higher median and maximum $\sigma_{M_s}$ values recorded. The $M_s = 2.17$ simulation still demonstrates the lowest error values, although the 0 - 20 spatial frequency removal case shows a median error exceeding 0.5 but remains below one. 
For the $M_s = 6.53$ and $M_s = 11.02$ simulations, the predictions reveal an almost inverse trend compared to other training scenarios, where the median and upper quartile $\sigma_{M_s}$ values decrease as the minimum retained spatial frequency increases. Despite this irregularity, the median error values for both simulations consistently remain below one. Across all three input map types and for all $M_s$ simulations, the error values stay below one—even when the largest amount of spatial frequencies is removed. This highlights the robustness of our CNN model in predicting $M_s$ using interferometric observations.

\section{Discussion}
\label{sec:discussion}
\subsection{Comparison with other methods}
In this study, we trained a CNN model to predict $M_s$ using spectroscopic observations. The CNN approach offers several advantages over traditional methods, which typically rely on the statistics of density fluctuations. The CNN model can be applied to velocity channel maps, suggesting the potential to map $M_s$ by using velocity information from spectroscopic observations. Additionally, the ability to accurately measure $M_s$ under conditions where lower spatial frequencies are removed is crucial for applications involving interferometric data. As demonstrated, our model performs well in noisy environments and when low spatial frequencies are unavailable. Furthermore, the CNN model is trained using numerical simulations. While turbulence simulations were employed in this study, more complex simulations may be required for additional physical processes. For example, when applying the CNN method to a turbulence-dominated molecular cloud, the effects of radiative transfer must be taken into account, and corresponding simulations should be incorporated into the CNN training. In molecular clouds with active star formation, additional factors such as gravitational collapse and outflow feedback should also be included to ensure accurate modeling.

\subsection{Measuring magnetic field strength in molecular cloud}
The magnetic field strength is crucial for understanding the role magnetic fields play in star formation \citep{1956MNRAS.116..503M,
10.1093/mnras/133.2.265, MK04,annurev:/content/journals/10.1146/annurev.astro.45.051806.110602,2012ARA&A..50...29C}. The CNN approach is particularly advantageous due to its flexibility in providing crucial parameters for measuring the magnetic field strength in molecular clouds. Previous research has demonstrated that CNN models can successfully predict the $M_A$ \citep{hu2023probing}. The two Mach numbers can be combined synergistically to estimate the magnetic field strength using the following relation \citep{lazarian2020obtainingmagneticfieldstrength, Hu_2023}:
\begin{equation}
\label{eq.B}
   B=c_{\mathrm{s}} \sqrt{4 \pi \rho} M_{\mathrm{s}} M_{\mathrm{A}}^{-1},
\end{equation}
where $B$ is the magnetic field strength, $c_s$ is the sound speed, and $\rho$ is the gas mass density. Thus, assuming $c_s$ is approximately constant in an isothermal molecular cloud, one could combine $M_A$ and $M_s$ to measure the spatial distribution of the total magnetic field strength. 

\subsection{Measuring 3D Galactic Magnetic Field strength}
Determination of the 3D Galactic magnetic field is vital to understanding many astrophysical processes and our own Galaxy, including the origin of ultra-high-energy cosmic rays \citep{PhysRevLett.89.281102, 10.3389/fspas.2022.900900,Lazarian_2023} and the modeling of foreground polarization for the CMB B-mode polarization detection \citep{PhysRevD.91.081303,  refId0}. 

Recently, research conducted has shown 21 cm neutral hydrogen thin channel maps can be used synergistically with the Galactic rotational curve to map the 3D GMF \citep{Lazarian_2018, 2023MNRAS.524.2994H,10.1093/mnras/stae146}. Intensity and velocity centroid maps are only able to map 2D POS magnetic fields. This makes channel input maps the ideal candidate for training CNN models to predict the 3D $M_s$ distribution, as well as the 3D GMF strength, based on Eq.~\ref{eq.B}. 

In addition, our CNN model demonstrated strong predictive capabilities, as evidenced by median $\sigma_{M_s}$ values consistently measured below 1 across all simulations when low spatial frequencies were removed. Given this success, magnetic field strength measurements could be extended extragalactically using high-resolution interferometric observation. 

\subsection{Understanding the compressibility of turbulence in ISM}
The information of $M_s$ is important in determining the gas compressibility $\beta=2(M_A/M_s)^2$. This compressibility is a key factor in understanding the fragmentation of molecular clouds, turbulent support, and the acceleration and diffusion of cosmic rays. In low-beta environments, where magnetic pressure dominates, magnetic fields play a significant role in shaping the gas dynamics. This can lead to higher gas compression in shocks and more efficient fragmentation of molecular clouds, facilitating the formation of dense structures and accelerating gravitational collapse, which increases the star formation efficiency \citep{1956MNRAS.116..503M,
10.1093/mnras/133.2.265, MK04}. On the other hand, the strong magnetic fields restrict cosmic ray motion in the direction perpendicular to the magnetic field, resulting in a smaller perpendicular diffusion coefficient \citep{2008ApJ...673..942Y,2013ApJ...779..140X,2022MNRAS.512.2111H}.

\section{Conclusion}
\label{sec:conclusion}

In conclusion, this study demonstrates the successful application of Convolutional Neural Networks (CNNs) in estimating the sonic Mach number $M_s$ of the ISM from spectroscopic data. The underlying physical principle of the CNN approach is that morphological changes in density and intensity structures are indicative of varying $M_s$ values. In highly supersonic media, characterized by significant small-scale density fluctuations, more pronounced small-scale filamentary structures are typically observed. By utilizing intensity, centroid, or velocity channel maps, the CNN effectively captures these structural changes, enabling accurate predictions of $M_s$ across diverse turbulence conditions, noise levels, and magnetic field inclinations. The CNN model's robustness, even under observational constraints such as the absence of low spatial frequencies, underscores its potential for analyzing interferometric data.
%

\begin{acknowledgments}
T.S. and A.L. acknowledge the support of NSF grants AST 2307840, and ALMA SOSPADA-016. Y.H. acknowledges the support for this work provided by NASA through the NASA Hubble Fellowship grant \# HST-HF2-51557.001 awarded by the Space Telescope Science Institute, which is operated by the Association of Universities for Research in Astronomy, Incorporated, under NASA contract NAS5-26555. This work used SDSC Expanse CPU, NCSA Delta CPU, and NCSA Delta GPU through allocations PHY230032, PHY230033, PHY230091, PHY230105, and PHY240183 from the Advanced Cyberinfrastructure Coordination Ecosystem: Services \& Support (ACCESS) program, which is supported by National Science Foundation grants \#2138259, \#2138286, \#2138307, \#2137603, and \#2138296. 
\end{acknowledgments}

\vspace{5mm}
\software{Python3 \citep{10.5555/1593511}; TensorFlow \citep{tensorflow2015-whitepaper}}


\newpage
\appendix
\section{High $M_A$ effect on $M_s$ prediction }
\label{appendix}
\begin{figure*}
    \centering \includegraphics[width=.75\linewidth]{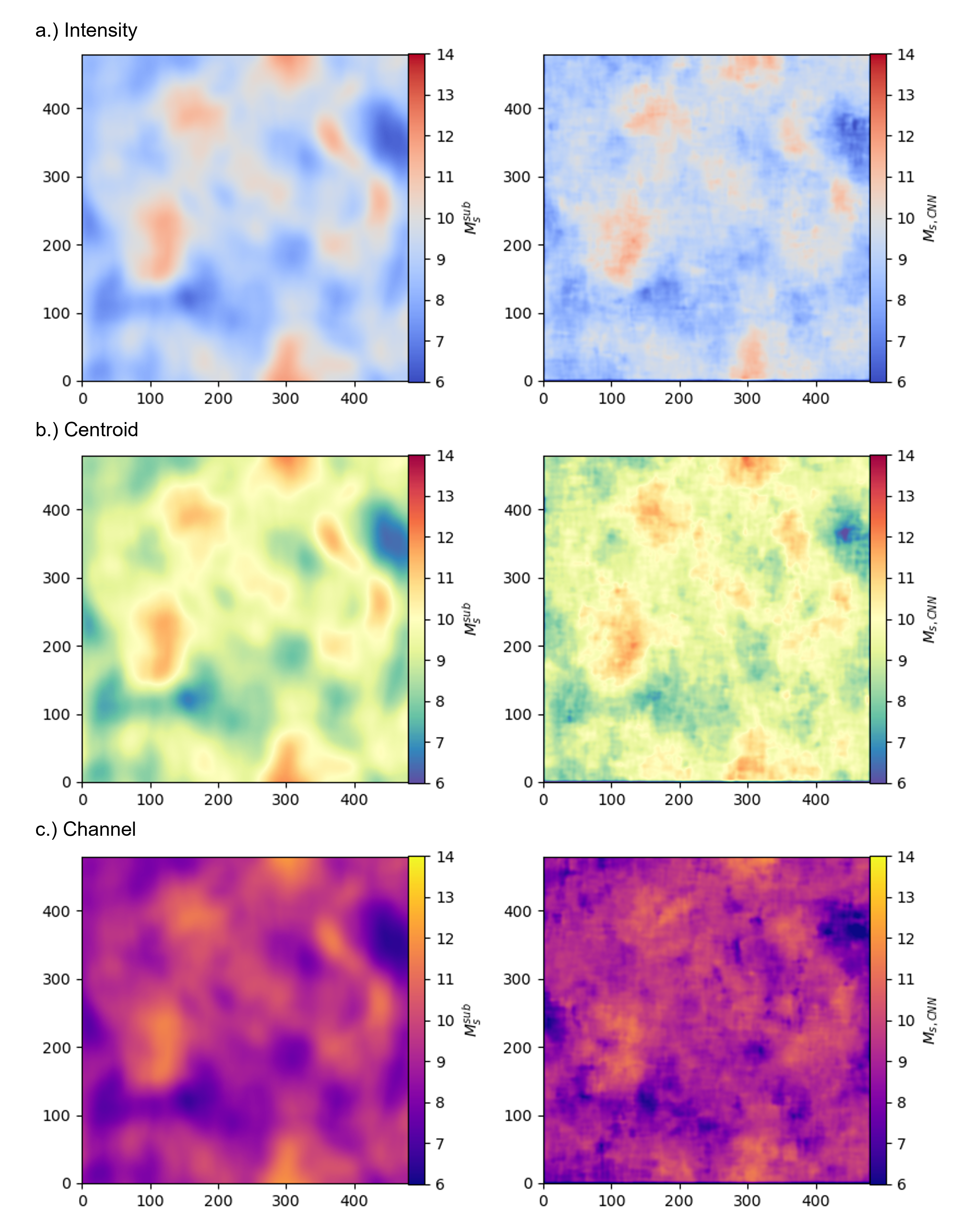}
    \caption{The right column presents a comparison of the CNN-predicted sonic Mach number $M_{s, {\rm CNN}}$ distribution from intensity (top), centroid (middle), and channel input maps (bottom) and the actual $M_{s}^{\rm sub}$ distribution in the left column. The supersonic simulation with $M_s=11.02$ and $M_A=1.07$ is used as an example here.}
    \label{fig:ms90 min05 and input maps 90 degrees prediction min05}
\end{figure*}

Fig.~\ref{fig:ms90 min05 and input maps 90 degrees prediction min05} displays the output of our CNN model when predicting $M_s$ for a simulation with an associated higher $M_A$, i.e., $M_A$ = 1.07, than the other sub-Alfv\'enic simulations used in this work. The prediction shows similar $M_s$ structures to the training map and demonstrates the robustness of our model to also train for varying $M_A$ number situations in supersonic molecular clouds.

\section{Test with unseen data }
\label{appendix2}
\begin{figure*}
    \centering \includegraphics[width=1.0\linewidth]{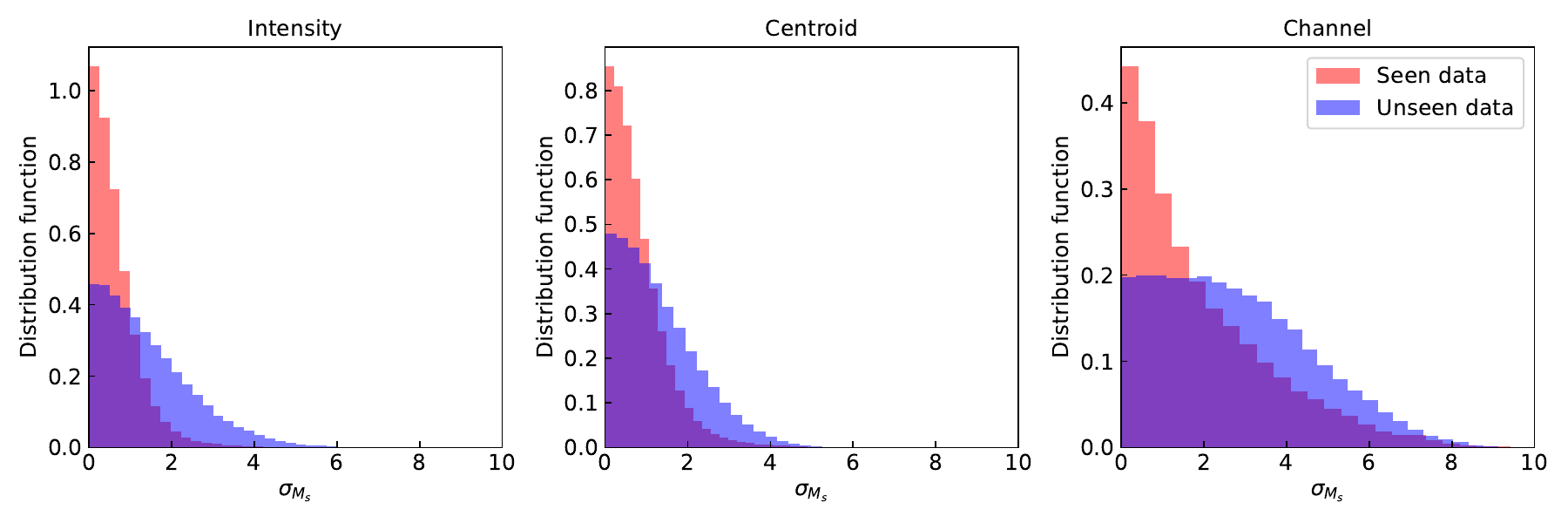}
    \caption{Histograms comparing the error $\sigma_{M_s}$ in predictions from simulation min06 ($M_s = 11.02$, $M_A = 0.43$) and from the unseen simulation min07 ($M_s = 10.50$, $M_A = 0.81$). }
    \label{fig:unseen}
\end{figure*}

Fig.~\ref{fig:unseen} presents a test of the CNN’s ability to predict $M_s$ for data not included in the training process. Specifically, we compare predictions from the simulation min06 ($M_s = 11.02$, $M_A = 0.43$), which was included in training, with predictions from a new simulation, min07 ($M_s = 10.50$, $M_A = 0.81$), which was not used during training and is treated as "unseen data." The histogram of $\sigma_{M_s}$ shows that the prediction error increases for unseen data.


\bibliography{sample631}{}
\bibliographystyle{aasjournal}



\end{document}